\documentstyle[12pt]{article}
\title{{Singularities in a Scalar Field Quantum Cosmology}
\thanks{This work is supported in part by funds provided by the U. S.
Department of Energy(D.O.E.) under contract \# DE-FC02-94ER40818 .}}
\author{\sc{Nivaldo A. Lemos}\thanks{On leave of absence from Departamento de
F\'{\i}sica, Universidade Federal Fluminense, Outeiro de S\~ao Jo\~ao
Batista s/n,\, 24020-005 Centro, Niter\'oi, RJ, Brasil.
Supported by Conselho Nacional de Desenvolvimento Cient\'{\i}fico e
Tecnol\'ogico(CNPq), Brasil.}\\
\small{\it Center for Theoretical Physics}\\
\small{\it Laboratory for Nuclear Science
and Department of Physics}\\
\small{\it Massachusetts Institute of Technology}\\
\small{\it Cambridge, MA 02139 - USA}\\
\small{e-mail: lemos@mitlns.mit.edu}
\\
\\
\\
\date{} }

\begin{document}
\pagestyle{myheadings}
\baselineskip 18pt

\maketitle
\begin{abstract}

The quantum theory of a spatially flat Friedmann-Robertson-Walker universe with
a massless scalar field as source is further
investigated. The classical model is
singular, and in the framework of a genuine canonical
quantization (Arnowitt-Deser-Misner formalism) a discussion is made of the
 cosmic evolution, particularly
of the quantum gravitational collapse problem. It is shown that
in a matter-time gauge such that time is identified with the scalar field the
classical model is singular either at $t=-\infty$ or at $t=+\infty$,
 but the quantum model is
nonsingular. The latter behavior disproves a conjecture according to which
quantum cosmological singularities are predetermined on the classical level
by the choice of time.
\vfill
\noindent CTP \# 2378 \hfill August 1995
\end{abstract}
\newpage

\noindent{\bf I. INTRODUCTION}
\vspace{0.75cm}

The problem of constructing a consistent quantum theory of the
gravitational field and its sources remains unsolved, in spite of great
efforts along several decades. Since the standard perturbative techniques
applied to quantum gravity appear to lead to a nonrenormalizable theory [1],
other lines of attack have been attempted. It is true that there has been
significant progress on nonperturbative canonical quantization of the full
gravitational field [2-5], but the enormous complexity of the problem calls for
manageable approximation schemes, one of the most attractive and fascinating of
which is quantum cosmology, initiated by  DeWitt [6] nearly thirty years
ago.

The essential idea of quantum cosmology [7] is to freeze out all but a finite
number of degrees of freedom of the system $-$ the gravitational field plus its
sources $-$ and then quantize the remaining ones. This procedure is known as
``minisuperspace quantization'', and although it cannot be strictly valid and
is open to criticism [8], it is expected to provide some general insights on
what an acceptable quantum theory of gravity should be like. This method has
been put to work for quantizing  Friedmann-Robertson-Walker(FRW) universes with
varying matter content such as a scalar field [9-11], a spinor field [12], dust
[13-16] or a Rarita-Schwinger field [17].

A fundamental issue of quantum cosmology is that of boundary or initial
conditions on the wave function of the universe [18], a subject that will not
be discussed here. Another outstanding problem is that of gravitational
collapse, or quantum cosmological singularities. On the classical domain the
celebrated theorems of Hawking and Penrose assert that singularities
inevitably occur in any spacetime obeying reasonable conditions on the causal
structure and matter content. At the quantum level the situation is not so
neat. The  canonical quantization method developed by Arnowitt, Deser
and Misner[19] seems to provide suitable means for studying quantum
cosmological
singularities. This approach consists in performing quantization in a reduced
phase space spanned by independent canonical variables, and demands a definite
choice of time. Although often leading to complicated and time-dependent
Hamiltonians, this formalism has the great advantage of reducing the problem
to one of standard quantum mechanics, enabling one to make full use of the
powerful theory of linear operators in Hilbert space. In so doing, at least for
FRW models one can define a quantum cosmological singularity with mathematical
precision and analyze in a satisfactorily rigorous fashion the influence of
quantum effects upon gravitational collapse.

It turns out that the issue of time in quantum cosmology(see [7] for references
in this connection) is entangled with the problem of quantum gravitational
collapse. Within the framework of the Arnowitt-Deser-Misner(ADM) genuine
canonical quantization, Gotay and Demaret [13] made a fairly general inquiry
into quantum cosmological singularities. They classify the time variable $t$ of
a classically singular model as either ``slow'', if the classical singularity
occurs at a finite value of $t$,  or ``fast'', if the classical singularity
occurs at $t=\pm\infty$. According to them, the existence of quantum
gravitational collapse is predetermined at the classical level by the choice of
time, the crucial distinction being between times that give rise to complete
or incomplete classical evolution.
Basing their contentions on their own findings concerning dust-filled FRW
models and on the models encountered in the literature until that date, they
summarized their analysis by conjecturing that ``self-adjoint quantum dynamics
in a slow-time gauge is always nonsingular'', whereas ``self-adjoint quantum
dynamics in a fast-time gauge is always singular''.

The first part of the above conjecture was disproved a few years ago by
exhibiting singular unitary [15] and strictly self-adjoint [16] quantum
cosmological models in a slow-time gauge. At that time no evidence was known
against the second part of the conjecture.

In the present paper we further study the quantum theory of a spatially flat
FRW
model with a massless scalar field as source, originally introduced by Blyth
and Isham [9]. We find that
for the choice of time
$t=\phi$, where $\phi$ is the scalar field, the classical model is singular
either at
$t=-\infty$ or at $t=+\infty$,    but the quantized model is
self-adjoint and nonsingular. Thus the second part of
the conjecture is  disproved.

This paper is organized as follows. In Section II the classical model is
specified and the solution to the equations of motion found originally in
[9] is reviewed. In Section III the ADM reduction of  phase space is
discussed for the choice of time referred to in the previous paragraph.
In Section IV
the model is quantized in the matter-time gauge    $t=\phi$ and shown to be
self-adjoint and free of singularity. Section V is
devoted to final remarks and a general
conclusion.

\vspace {1.25cm}

\noindent{\bf II. DESCRIPTION OF THE CLASSICAL MODEL}

\vspace {.75cm}
We shall be interested in homogeneous and isotropic universes described by the
FRW metrics

$$ds^2 = g_{\mu\nu} dx^{\mu} dx^{\nu} = - N(t)^2 dt^2
+ R(t)^2 {\sigma}_{ij} dx^i dx^j
\,\,\, , \eqno(1)$$
\\
\noindent where ${\sigma}_{ij}$ is the   metric for a 3-space
of constant curvature
$k= +1, 0$ or $-1$, corresponding to spherical, flat or hyperbolic spacelike
sections, respectively.

The classical action  (in units such that $c=16\pi G=1$) is

$$ S= - \int_M\,d^4x \sqrt{-g}\,\, ^{(4)}R \,
-\, 2 \int_{\partial M}\,d^3x\sqrt{h}\, h_{ij} K^{ij} \, + \,
\frac{1}{2}  \int_M\,d^4x  \sqrt{-g}\,{\partial}_{\mu}\phi
{\partial}^{\mu}\phi \,\,\, \eqno(2)$$
\\

\noindent where $\phi$ is a massless scalar field, $^{(4)}R$ is the scalar
curvature derived from the spacetime metric $g_{\mu\nu}$, $h_{ij}$ is the
metric on the boundary $\partial M$, and $K^{ij}$ is the second fundamental
form of the boundary [20]. The surface term is necessary in the
path-integral formulation of quantum gravity in order to rid the
Einstein-Hilbert Lagrangian of second-order derivatives. Compatibility with
the homogeneous spacetime metric requires a space-independent scalar field,
that is, $\phi = \phi (t)$.

In the geometry characterized by (1) the appropriate boundary condition for the
action principle is to fix the initial and final hypersurfaces of constant
time. The second fundamental form of the boundary becomes
$K_{ij} = - {\dot{h}}_{ij}/2N$. From now on an overall factor of the spatial
integral of $(det \sigma)^{1/2}$ will be discarded, since it has no effect on
the equations of motion. Insertion of the metric (1) and of the homogeneous
scalar field into (2) yields the reduced action

$$S_r = \int dt\, L \eqno(3)$$
\\
\noindent with the Lagrangian

$$ L =  \frac{6R}{N}{\dot R}^2 - 6kNR - \frac{1}{2}\frac{R^3}{N}{\dot \phi}^2
\,\,\, . \eqno(4)$$
\\
\noindent The canonical momentum conjugate to $R$ is

$$p_R = \frac{\partial L}{\partial {\dot R}} =  12\frac{R{\dot R}}{N}
\eqno(5)$$
\\
\noindent whereas the momentum conjugate to $\phi$ is

$$p_{\phi} = \frac{\partial L}{\partial {\dot \phi}} =
- \frac{R^3}{N}{\dot \phi} \,\,\, , \eqno(6)$$
\\

\noindent so that the classical action can be recast in the Hamiltonian form

$$ S_r = \int dt \Biggl \{ {\dot R}p_R + {\dot \phi}p_{\phi}
- N \Bigl ( \frac{p_R^2}{24R} - \frac{p_{\phi}^2}{2R^3} + 6kR \Bigr )
\Biggr \} \,\,\, . \eqno(7)$$
\\

If one inserts the metric (1) and the homogeneous scalar field $\phi (t)$ into
the field equations derived from the full action (2), the resulting equations
of motion are identical to those that follow from the reduced action (7)
under variation of $R$, $\phi$ and $N$. These
classical equations of motion were explicitly solved in [9] for closed or open
models. For the purpose of quantization
we shall direct our attention only to the simplest case $k=0$, for which
Einstein's  ``$G_{00}$ equation''  is

$$3\frac{{\dot{R}}^2}{R^2} = \frac{1}{4}{{\dot{\phi}}^2 } \,\,\, .\eqno(8)$$
\\
\noindent In the gauge $t=\phi$ the above equation is equivalent to

$$\dot{R} =   \left\{ \begin{array}{cl}
              R/{\sqrt {12}} & \mbox{if ${\dot R}>0$} \\
              \\
              - R/{\sqrt {12}} & \mbox{if ${\dot R}<0$}
              \end{array} \right.   \,\,\, .    \eqno(9)$$
\\
\\
\noindent The  field equations allow of expanding or contracting universes,
that is, $R(t) = R_0 \exp (\pm t/{\sqrt{12}})$, where $R_0$ is an arbitrary
positive constant. These are mutually exclusive solutions, depending on the
initial conditions. The model is singular at  $t=-\infty$ in the expanding case
or at  $t=+\infty$ in the contracting case. As will be seen, although
classically
the existence of one of these solutions automatically precludes the existence
of the other, at the quantum level they not only coexist but also interfere
with each other.

The form (7) of the reduced action shows clearly that the lapse function $N$
plays the role of a Lagrange multiplier. Variation with respect to
$N$ leads to the
super-Hamiltonian constraint

$$ \frac{p_R^2}{24R} - \frac{p_{\phi}^2}{2R^3} + 6kR = 0 \,\,\, , \eqno(10)$$
\\

\noindent which for $k=0$ and with the use of (5) and (6) is easily seen to be
identical to Eq.(8).
This constraint reveals that the phase space $(R,\phi ,p_R,p_{\phi})$
is too large, so that a bona fide canonical quantization can only be performed
after going over to a reduced phase space spanned by independent canonical
variables alone. This can be achieved         by first making a choice of time
and then solving the constraint equation (9) for the canonical variable
conjugate to the time chosen in the first step. This ensures that the final
action preserves its canonical form, with a Hamiltonian identical to the
variable whose Poisson bracket is unity with  whatever was chosen as time,
but now expressed as a
function of the remaining independent canonical variables [19]. This is the
essence                                                 of the ADM formalism,
which will be illustrated below for an specific choice of time.

\vskip 1.25cm

\noindent {\bf III. MATTER-TIME GAUGE AND ADM REDUCTION}

\vskip .75cm

For the sake of simplicity, from now on our attention will be focussed solely
on the spatially flat case, that is,
$k=0$. Let us make the choice of time  $t=\phi$, the matter field
itself providing a clock by means of which
the evolution of the system can be followed. According to the ADM prescription,
the Hamiltonian in the reduced phase space is $H = - p_{\phi}$. Now,
solving Eq.(10) for $p_{\phi}$  and picking up the
negative square-root gives rise to the reduced Hamiltonian

$$ H = - p_{\phi} = \frac{1}{\sqrt{12}}\, R\, \vert p_R \vert
\,\,\, . \eqno(11)$$
\\

\noindent It is important to notice that in the gauge $t=\phi$
it follows from Eq.(6) that $p_{\phi}<0$ since $R>0$ and $N>0$
{\it by definition}. This is the reason why the positive solution for
$p_{\phi}$ was abandoned. One sees, therefore, that the Hamiltonian
(11) is positive. Hamilton's equation of motion  for $R$
in the reduced phase space  is

$${\dot R} = \frac{\partial H}{\partial p_R} =
              \left\{ \begin{array}{cl}
              R/{\sqrt {12}} & \mbox{if $p_R > 0$} \\
              \\
              - R/{\sqrt {12}} & \mbox{if $p_R < 0$}
              \end{array} \right. \,\,\, .      \eqno(12)$$
\\
\\
\noindent Because $R>0$ and $N>0$  by definition, it is a consequence of Eq.(5)
that $p_R$ and $\dot{R}$ have the same sign, so that Eqs.(9) and (12)
are identical.
This completes the verification that the equations of motion generated by the
reduced Hamiltonian (11) are the same as those that arise from
variation of the  action (7) in the extended phase space.

The reduced phase space ${\cal P} = (R,p_R)$ is the union
${\cal P} = {\cal P}_+ \cup {\cal P}_- $ of the two disjoint sets
${\cal P}_+ = (0,\infty ) \times (0,\infty )$ and
${\cal P}_- = (0,\infty ) \times (-\infty ,0)$. From
Eq.(8) in the gauge $t=\phi$ it follows that $\dot R$ can
never vanish, so that the line $p_R =0$
does not belong to the reduced phase space.
The sets
${\cal P}_+$ and ${\cal P}_-$ are disconnected in the sense that the dynamical
trajectories remain entirely confined to one of them,
selected according to the initial conditions,
and cannot cross the border $p_R = 0$ between them.

As remarked previously, for  the present choice of time
the scale factor vanishes
and Riemann tensor invariants such as $^{(4)}R$ become infinite
either when $t= -\infty$ or when $t= +\infty$.
Therefore the classical model is singular and
$t=\phi$ is a ``fast'' time in accordance with the
terminology introduced in [13].

\vskip 1.25cm

\noindent {\bf IV. QUANTIZATION IN THE MATTER-TIME GAUGE }

\vskip .75cm

As discussed above, in the gauge $t=\phi$ the classical Hamiltonian function
is (11). An operator corresponding to $R\vert p_R \vert$ can be naturally
defined as the
positive square root of an operator corresponding to $R^2 p_R^2$. Thus, we look
for a positive Hamiltonian operator whose square has as classical counterpart
the square of the Hamiltonian function (11).
Following Blyth and Isham [9],
such a  positive self-adjoint Hamiltonian can be
constructed as the square root of the positive self-adjoint operator

$${\hat {\cal O}} = - \frac{1}{12} R^{\nu} \frac{d}{dR}
R^{2-2\nu}\frac{d}{dR} R^{\nu} \eqno(13)$$
\\

\noindent with a suitable domain of definition, where the parameter $\nu$
reflects factor-ordering ambiguities.
In Ref.[9] the choice $\nu = 0$ was
made, but it turns out, as will be shown below, that that there is a better
choice of the parameter $\nu$ that makes easier the analysis of the quantum
dynamics.
Therefore we take

$${\hat H}^2 =  - \frac{1}{12} R^{\nu} \frac{d}{dR}
R^{2-2\nu}\frac{d}{dR} R^{\nu} = - \frac{1}{12}
\Bigl [  \frac{d}{dR}
R^{2}\frac{d}{dR}  + \nu (1 - \nu) \Bigr ]  \eqno(14)$$
\\

\noindent acting on $L^2(0,\infty )$. A great deal of simplification is
achieved by means of the unitary mapping from ${\cal H}=L^2(0,\infty )$
onto ${\tilde{\cal H}} = L^2(-\infty ,\infty )$ defined by [9]

$$\tilde{\psi}(y) = e^{-y/2} \psi (e^{-y}) \,\,\, , \eqno(15)$$
\\

\noindent which is tantamount to the change of variable $R=e^{-y}$. Indeed,
the expectation value

$$\langle {\hat R} \rangle _{\psi} =
\langle \psi \vert  {\hat R} \vert \psi \rangle  =
\int_0^{\infty}\, R \, \vert \psi (R) \vert ^2 \, dR
 \eqno(16)$$

\noindent becomes

$$ \langle {\hat R} \rangle _{\psi} =
\int_{-\infty}^{\infty}\, e^{-y}\,  \vert \psi (e^{-y}) \vert ^2 \,
e^{-y}\, dy =
\int_{-\infty}^{\infty}\, e^{-y}\,  \vert \tilde{\psi} (y) \vert ^2 \, dy
=    \langle e^{-\hat y} \rangle _{\tilde{\psi}} \,\,\, .
 \eqno(17)$$
\\

\noindent The transformed Hamiltonian squared is easily obtained by demanding
that its expectation value in a state $\tilde{\psi} \in L^2(-\infty ,\infty )$
be equal to the expectation value of (14) calculated in the state
$\psi \in L^2(0,\infty )$ with  $\psi$ and  $\tilde{\psi} $ related by Eq.(15).
The result is

$${\hat{\tilde{H}}}^2 = \frac{1}{12} \Bigl [ -\frac{d^2}{dy^2} + \frac{1}{4} +
\nu (\nu -1) \Bigr ] \,\,\, , \eqno(18)$$
\\

\noindent which, with the choice $\nu = 1/2$, reduces to the simple form

$${\hat{\tilde{H}}}^2 = - \frac{1}{12} \frac{d^2}{dy^2} \,\,\, . \eqno(19)$$
\\

It is very convenient to investigate the quantum dynamics in the momentum
representation  in the transformed Hilbert space
$\tilde{\cal H}$.  Then $-id/dy$ becomes the operator of
multiplication by $p$ and the positive square root of (19) is such that

$$({\hat{\tilde{H}}}\tilde{\psi})(p) =  \frac{1}{\sqrt{12}} \vert p \vert
\tilde{\psi}(p) \eqno(20)$$
\\

\noindent on the dense domain

$${\cal D} = \Bigl \{\tilde{\psi} \in L^2(-\infty,\infty ) \,
\vert \int_{-\infty}^{\infty} p^2
\vert \tilde{\psi} (p) \vert ^2 \, dp \, < \infty \Bigr \} \,\,\, . \eqno(21)$$
\\

\noindent Given
 an initial wave function  ${\tilde{\psi}}_0(p)$    at $t=t_0$,
one finds that at time $t$

$$\tilde{\psi}(p,t) =
\Bigl ( exp[-i(t-t_0){\hat{\tilde{ H }}}]{\tilde{\psi}}_0 \Bigr ) (p)
= e^{-i(t-t_0)\vert p \vert /\sqrt{12}}\, {\tilde{\psi}}_0(p)
\,\,\, . \eqno(22)$$
\\

The singularity criterion to be adopted here is the following [13,21]: the
quantum system is singular at a certain instant if $\langle \psi \vert
{\hat f} \vert \psi \rangle = 0$ for any quantum observable $\hat f$ whose
classical counterpart $f$ vanishes at the classical singularity, $\psi$ being
any state of the system at the instant under consideration. For  models
of the FRW type the
relevant quantum observable is ${\hat f} = \hat R$, since $R=0$
defines the classical singularity. This criterion is in agreement with the
usage in quantum cosmology. Indeed, since $\hat R$ is a positive operator
on $L^2(0,\infty )$, if $\langle {\hat R} \rangle _t = 0$ then $\psi (t)$
is sharply peaked at $R=0$, and a strong peak in the wave function at a certain
classical configuration is regarded in quantum cosmology as a prediction of the
occurrence of such a configuration [7].

Accordingly, if $\langle \psi (t)\vert  {\hat R} \vert \psi (t)\rangle $ never
vanishes for some evolving state $ \psi (t)  $ then the model is
nonsingular. Let us take as initial state the Gaussian wave packet

$${\tilde{\psi}}_0(p) = {\pi}^{-1/4} e^{-p^2/2} \,\,\, ,   \eqno(23) $$
\\

\noindent from which one finds

$$ \tilde{\psi}(y,t) = \frac{1}{\sqrt{2\pi}}  \int_{-\infty}^{\infty}
\tilde{\psi}(p,t)e^{ipy}dp =
\frac{{\pi}^{-1/4}}{\sqrt{2\pi}}
\int_{-\infty}^{\infty} \exp\Bigl [ - \frac{i}{\sqrt{12}}(t-t_0)\vert p \vert
- \frac{p^2}{2} + ipy \Bigr ]\, dp \,\,\, . \eqno(24)$$
\\

\noindent In terms of the convenient quantities

$$ {\xi}_{\pm}(y,t) = y \pm \frac{t-t_0}{\sqrt{12}} \,\,\, \eqno(25)$$
\\

\noindent one can reexpress Eq.(24) as

$$  \tilde{\psi}(y,t) = \frac{{\pi}^{-1/4}}{\sqrt{2\pi}} \Biggl \{
\int_0^{\infty}\, \cos (p {\xi}_+(y,t))e^{-p^2/2} dp -
i\int_0^{\infty}\, \sin (p {\xi}_+(y,t))e^{-p^2/2} dp $$

$$+ \int_0^{\infty}\, \cos (p {\xi}_-(y,t))e^{-p^2/2} dp +
i\int_0^{\infty}\, \sin (p {\xi}_-(y,t))e^{-p^2/2} dp  \Biggr \} \,\,\, .
\eqno(26)$$
\\

\noindent These integrals can be explicitly evaluated to yield [22]

$$  \tilde{\psi}(y,t) = \frac{{\pi}^{-1/4}}{\sqrt{2\pi}} \Biggl \{
\frac{\sqrt{2\pi}}{2} e^{-{\xi}_+(y,t)^2/2} -
i{\xi}_+(y,t)\, _1F_1\Bigl ( 1,\frac{3}{2};-\frac{{\xi}_+(y,t)^2}{2} \Bigr ) +
\bigl ( \xi _+ \leftrightarrow \xi _- \bigr )^* \Biggr \} \,\,\, , \eqno(27)$$
\\

\noindent where $_1F_1$ denotes a degenerate (confluent) hypergeometric
function and the asterisk stands for complex conjugate.
The above wave function is the superposition of two wave packets,
one centered on $y = - (t-t_0)/\sqrt{12}$ and the other on
 $y = + (t-t_0)/\sqrt{12}$. The first packet corresponds to an expanding
universe, while the second one describes a contracting universe.

The initial expectation value of $\hat{R} $ is finite and can be computed once
${\tilde{\psi}}_0(y)$ has been found. We have

$${\tilde{\psi}}_0(y) = \frac{1}{\sqrt{2\pi}}
\int_{-\infty}^{\infty}\,{\tilde{\psi}}_0(p)e^{ipy}\, dy =
{\pi}^{-1/4}\, e^{-y^2/2} \,\,\, , \eqno(28)$$
\\

\noindent hence

$$ \langle {\hat R} \rangle _{t_0} = \int_{-\infty}^{\infty}
e^{-y}\, {\pi}^{-1/2}\, e^{-y^2}\, dy = e^{1/4} \,\,\, . \eqno(29)$$
\\

\noindent The general structure of Eq.(27) is

$$\tilde{\psi}(y,t) = \tilde{\psi}_1(\xi _+) - i\tilde{\psi}_2(\xi _+)
+ \tilde{\psi}_1(\xi _-) + i\tilde{\psi}_2(\xi _-)  \eqno(30)$$
\\
\noindent with $\tilde{\psi}_1\, , \,  \tilde{\psi}_2 $ real functions, and, in
particular,

$$\tilde{\psi}_1(x) = \frac{{\pi}^{-1/4}}{2} e^{-x^2/2} \,\,\, . \eqno(31)$$
\\

\noindent Therefore, since   $\tilde{\psi}_1$ is a positive function,

$$ \vert \tilde{\psi}(y,t) \vert ^2 \geq
\vert \tilde{\psi}_1(\xi _+) +  \tilde{\psi}_1(\xi _-) \vert ^2 \geq
\vert \tilde{\psi}_1(\xi _+) \vert ^2
 + \vert \tilde{\psi}_1(\xi _-) \vert ^2 \,\,\, \eqno(32)$$
\\

\noindent whence

$$ \langle {\hat R} \rangle _{t} \geq
\int_{-\infty}^{\infty}\, e^{-y}\,
 \vert \tilde{\psi}_1(\xi _+(y,t)) \vert ^2 \, dy
+ \int_{-\infty}^{\infty}\, e^{-y}\,
\vert \tilde{\psi}_1(\xi _-(y,t)) \vert ^2 \, dy \,\,\, . \eqno(33)$$
\\

\noindent A straightforward evaluation of the above
integrals with the help of (25)
and (31) furnishes

$$ \langle {\hat R} \rangle _{t} \geq \frac{\langle {\hat R} \rangle _{t_0}}{2}
\cosh\Bigl ( \frac{t-t_0}{\sqrt{12}} \Bigr ) \geq
\frac{\langle {\hat R} \rangle _{t_0}}{2}
\,\,\, . \eqno(34)$$
\\

\noindent It is thus established that the expectation value
$\langle {\hat R} \rangle _{t}$
never vanishes, and, in particular, $\langle {\hat R} \rangle _{t}$
tends to infinity as $t \rightarrow \pm \infty$ (classical singularity). This
constitutes an example of a nonsingular self-adjoint quantum cosmological
model in a fast-time gauge, and allows us to  conclude that
the second part of the conjecture advanced by Gotay and Demaret [13] is not
true. We remark that the special choice $\nu = 1/2$ is not a weak point of our
argument. The conjecture asserts that {\it all} self-adjoint quantum
cosmological models in a fast-time gauge are singular. Here a particular
counterexample (with $\nu = 1/2$) has been exhibited of a nonsingular
self-adjoint quantum cosmological model in a fast-time gauge.


\vskip 1.25cm


\noindent {\large{\bf V. CONCLUSION}}

\vskip .75cm

The main finding of the present work is that, contrary to a plausible belief,
quantum cosmological models in fast-time gauges are not necessarily singular.
Combined with the results obtained in [15] and [16],
our present investigation reveals
that the occurrence of  gravitational collapse at the quantum level
 is not classically
predetermined by the choice of a ``fast'' or ``slow'' time, such a
classification not being very relevant      to the problem of quantum
gravitational collapse.  It thus appears that the issue of time in quantum
cosmology and quantum gravity is actually deeper and more complicated than was
guessed hitherto. The apparent absence of an intrinsic time variable in the
general theory of relativity, and the physical inequivalence of different
choices of time in quantum cosmology remain as challenges to be met by any
candidate to a viable quantum theory of the gravitational field.

\vskip 1.25cm

\noindent {\large{\bf ACKNOWLEDGMENT}}

\vskip .75cm

The author is grateful to J. A. S. Lima for useful suggestions and for
reading the manuscript.
\newpage

\centerline{\bf REFERENCES}
\begin{description}

\item{[1]} G. 't Hooft in {\it Recent Developments in Gravitation, Carg\`ese},
           ed. M. L\'evy and S. Deser (Plenum, New York,1979); G. 't Hooft and
           M. Veltman, Ann. Inst. Henri Poincar\'e {\bf 20}, 69 (1974); M. H.
           Goroff and A. Sagnotti, Nucl. Phys. {\bf B266}, 709 (1986).

\item{[2]} A. Ashtekar, Phys. Rev. Lett. {\bf 57}, 2244 (1986); Phys. Rev.
           {\bf D36}, 1587 (1987).

\item{[3]} T. Christodoulakis and J. Zanelli, Class. Quantum Grav. {\bf 4},
           851 (1987).

\item{[4]} T. Jacobson and L. Smolin, Nucl. Phys. {\bf B299}, 295 (1988).

\item{[5]} T. Fukuyama and K. Kamimura, Phys. Rev. {\bf D41}, 1105 (1988).

\item{[6]} B. S. DeWitt, Phys. Rev. {\bf 160}, 1113 (1967).

\item{[7]} For a guide to the literature on quantum cosmology, see J. A.
           Halliwell in {\it Quantum Cosmology and Baby Universes}, ed. S.
           Coleman, J. B. Hartle, T. Piran and S. Weinberg (World Scientific,
           Singapore,1991).

\item{[8]} K. V. Kucha\v{r} and M. P. Ryan Jr., Phys. Rev. {\bf D40}, 3982
           (1989).

\item{[9]} W. F. Blyth and C. J. Isham, Phys. Rev. {\bf D11}, 768 (1975).

\item{[10]} N. A. Lemos, Phys. Rev. {\bf D36}, 2364 (1987).

\item{[11]} X. Hu and Y. L. Wu, Phys. Lett. {\bf A125}, 362 (1987).

\item{[12]} C. J. Isham and J. Nelson, Phys. Rev. {\bf D10}, 3226 (1974).

\item{[13]} M. J. Gotay and J. Demaret, Phys. Rev. {\bf D28}, 2402 (1983).

\item{[14]} P. S. Joshi and S. S. Joshi, Mod. Phys. Lett. {\bf A2},
            913 (1987).

\item{[15]} N. A. Lemos, Phys. Rev. {\bf D41}, 1358 (1990).

\item{[16]} N. A. Lemos, Class. Quantum Grav. {\bf 8}, 1303 (1991).

\item{[17]} T. Christodoulakis and C. G. Papadopoulos, Phys. Rev. {\bf D38},
            1063 (1988).

\item{[18]} A. Vilenkin, Phys. Rev. {\bf D50}, 2581 (1994); for a review of
            this subject see also [7], where references can be found to the
            seminal work of Hawking, Hartle, Vilenkin and Linde.

\item{[19]} R. Arnowitt, S. Deser and C. W. Misner in {\it Gravitation, an
            Introduction to Current Research}, ed. L. Witten (Wiley,New
            York,1962).

\item{[20]} See, for example, S. W. Hawking in {\it Quantum Gravity and
            Cosmology}, ed. H. Sato and T. Inami (World Scientific,
            Singapore,1986).

\item{[21]} F. Lund, Phys. Rev. {\bf D8}, 3253 (1973); M. J. Gotay and J. A.
            Isenberg, Phys. Rev. {\bf D22}, 235 (1980); J. A. Isenberg and M.
            J. Gotay, Gen. Rel. Grav. {\bf 13}, 301 (1981).

\item{[22]} I. S. Gradshteyn and I. M. Ryzhik, {\it Tables of Integrals,
            Series, and Products} (Academic, New York,1980), {\it Corrected and
            Enlarged Edition}, p.480

\end{description}
\end{document}